\documentclass[usenatbib,usegraphicx]{mn2e}
\usepackage{amssymb}

\DeclareGraphicsExtensions{.eps,.ps}

\newcommand{\upf}[1]{\raisebox{0.5ex}[0pt]{#1}}

\title[The embedded cluster or association Trumpler\,37 in IC\,1396]{The embedded cluster or association Trumpler\,37 in IC\,1396: a search for evolutionary constraints}

\author[T.A. Saurin, E. Bica and C. Bonatto]{T.A. Saurin$^{1}$\thanks{E-mail: tiago.saurin@ufrgs.br (TAS)},
E. Bica$^{1}$\thanks{E-mail: bica@if.ufrgs.br (EB)} and C. Bonatto$^{1}$\thanks{E-mail: charles@if.ufrgs.br (CB)}\\
$^{1}$Universidade Federal do Rio Grande do Sul, Departamento de Astronomia, CP\,15051, RS, Porto Alegre 91501-970, Brazil}

\begin{document}

\date{Accepted 2012 January 12. Received 2012 January 11; in original form 2011 September 15}

\pagerange{\pageref{firstpage}--\pageref{lastpage}} \pubyear{2011}

\maketitle

\label{firstpage}

\begin{abstract}
It is currently widely accepted that open star clusters and stellar associations result from the evolution of embedded star clusters.
Parameters such star formation efficiency, time-scale of gas removal and velocity dispersion can be determinants of their future as bound or unbound systems.
Finding objects at an intermediate evolution state can provide constraints to model the embedded cluster evolution.
In the H\,II region IC\,1396, Trumpler\,37 is an extended young cluster that presents characteristics of an association.
We employed the Two Micron All Sky Survey (2MASS) photometry to analysing its structure and stellar content, and determining its astrophysical parameters.
We also analysed 11 bright-rimmed clouds in IC\,1396 in order to search for young infrared star clusters, and the background open star cluster Teutsch\,74, to verify whether it has any contribution to the observed stellar density profile of Trumpler\,37.
The derived parameters and comparison with template objects from other studies lead us to conclude that Trumpler\,37, rather than as a star cluster, will probably emerge from its molecular cloud as an OB association.
\end{abstract}

\begin{keywords}
open clusters and associations: individual: Trumpler\,37 --
open clusters and associations: individual: Teutsch\,74
\end{keywords}

\section{Introduction}
\label{sec:in}
Studies comparing embedded and open star cluster frequencies have found that the
former type is more common (e.g. \citealt{lada03,bonbic11}). The drop
observed in the latter frequency has been interpreted as a result of
{\it infant mortality}. The paradigm is that most young clusters dissolve
in the field because they cannot return to dynamical equilibrium after the
residual gas expulsion (\citealt{tutuko78,felkro05,basgoo06}).

A star formation efficiency (SFE -- the fraction of molecular cloud gas
converted into stars) higher than 50\,per cent
is pointed out as the main factor to determine whether a star cluster will
survive an abrupt gas removal \citep{hills80}. However, the observed SFE
has been always less than 50\,per cent
(\citealt{lada03}, and references therein). Furthermore, \citet{kroupa05} showed
that a single O-type star may inject, by means of winds and electromagnetic
radiation, an energy higher than the cluster binding energy. This occurs on a
time-scale shorter than the cluster crossing time ($\tau_{cr}$ -- the time-scale
over which the system as a whole will respond to changes in the overall
potential). Therefore, other parameters such as the time-scale of gas removal,
tidal radius and initial cluster distribution function must play some role in
this picture \citep{boikru03}. A low stellar velocity dispersion immediately
before the onset of gas removal can be more important than the SFE for the
cluster survival \citep{good09}. In addition, even a cluster with low SFE
($\sim$20\,per cent)
might survive gas removal if it has a complex structure of subclusters.
Simulations by \citet{kruijs11} have pointed out that the subclusters in an
embedded cluster are close to virial equilibrium, and that they are weakly
affected by gas removal. In this case, young clusters survive to
{\it infant mortality} as bound systems. The merging of subclusters may form a
more massive object with a much higher escape velocity \citep{felkro05}.
Therefore, the low efficiency formation of clusters would be the result of
another disruptive process, the tidal shocks from the dense star-forming
regions. Only clusters that migrate out of their natal environment in a short
enough time-scale should survive.

The disruptive processes (residual gas expulsion and tidal shocks by the
surrounding clouds) cause an abrupt change in the gravitational potential that
may lead an embedded star cluster to endure heavy stellar loss and expand to
become a stellar association. However, the stars in these unbound systems are
expected to keep a common proper motion by tens of millions of years
\citep{basgoo06}. Hence, distinction between star clusters (bound systems)
and stellar associations may be difficult to establish for distant objects
because of limitations to observe the proper motions of individual stars.
Therefore, \citet*{giepor11} proposed to use the ratio between stellar age and
crossing time as a criterion to distinction. Thus, unbound systems would have
Age/$\tau_{cr}$\,$<$\,1 and bound systems, Age/$\tau_{cr}$\,$>$\,1.

Examples of possible intermediate objects between embedded star clusters and
stellar associations have been searched in order to determine constraints to the
evolution of the embedded star clusters
(e.g. \citealt*{saurin10,bonbic10,bonbic11}). Evidence of dissolving clusters
with escaping stars can be provided by the analysis of extensions in radial
profiles for large radii reflecting changes in the gravitational potential
\citep{basgoo06}.

In the present paper we analyze Trumpler\,37 (Collinder\,439) using the
Two Micron All Sky Survey
(2MASS\footnote{\it http://www.ipac.caltech.edu/2mass}; \citealt{skruts06})
$J$, $H$ and $K_S$ photometry and determine some astrophysical parameters that
characterize this object that might be evolving to an OB association.

\begin{figure*}
\resizebox{\hsize}{!}{\includegraphics{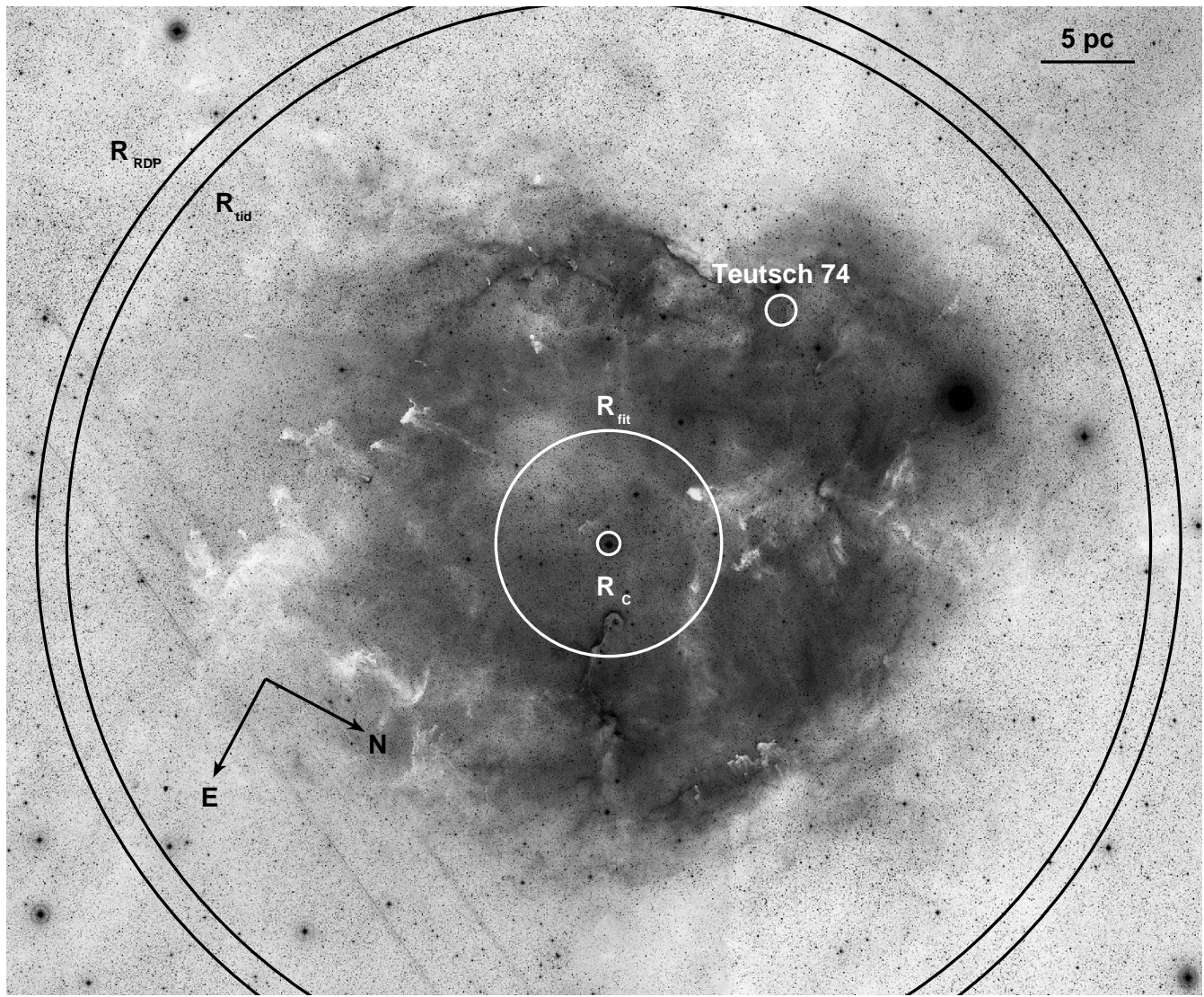}}
\caption[]{Combined B and R bands from DSS images with dimensions
4.3$^{\circ}\times$3.6$^{\circ}$ of Trumpler\,37 embedded in IC\,1396. Circles
of radii {\it R$_{c}$} (Table\,\ref{tab:tr37p}), {\it R$_{fit}$},
{\it R$_{tid}$} and {\it R$_{RDP}$} (Table\,\ref{tab:tr37r}) used to analyse
Trumpler\,37 are indicated. The background open cluster Teutsch\,74
(Sect.\,\ref{sec:teu74}) is encircled in the upper right.}
\label{fig:ic1396}
\end{figure*}

Trumpler\,37 is embedded in the H\,II region IC\,1396 -- catalogued by
\citet{sharpl59} as Sh2-131 -- a cloud located close to the Cep\,OB2
association. Fig.\,\ref{fig:ic1396} shows a Digitized Sky Survey
(DSS\footnote{\it http://cadcwww.dao.nrc.ca/dss/}; \citealt{reid91}) image of
the complex.
The brightest star in the complex is HR\,8281 (HD\,206267), a spectroscopic
binary of O6.5V((f))+O9:V types that belongs to the multiple system ADS\,15184
\citep{tokovi97}. This complex also includes many dark nebulae and
substructures, remarkably the Elephant Trunk Nebula (IC\,1396A) with many young
stellar objects (e.g. \citealt{reach04,mercer09}). In a study by
\citet{weikard96} about the bright-rimmed clouds in IC\,1396 it was estimated a
gas and dust mass {\it M$_{cld}$}\,=\,12$\times$10$^3$M$_{\odot}$ for the whole
complex. There is some spread in the distance estimates for
Trumpler\,37/IC\,1396, e.g. 705\,pc \citep{becfen71}; 860\,pc \citep*{blfist82};
798\,pc \citep{batdol91} and 835\,pc \citep{kharch05}. These values imply a
mean {\it d$_{\odot}$}\,=\,800$\pm$60\,pc, which we adopt hereafter.

This paper is organised as follows. In Sect.\,\ref{sec:cmd} we build
colour-magnitude diagrams (CMDs) and a colour-colour diagram to estimate some
astrophysical parameters of Trumpler\,37. In Sect.\,\ref{sec:rdp} we analyze the
radial density profiles (RDPs) and derive structural parameters. In
Sect.\,\ref{sec:dis} we discuss the results and use diagnostic diagrams to
compare the Trumpler\,37 parameters with those of a template. In
Sect.\,\ref{sec:con} we provide the concluding remarks of this work.

\section{2MASS photometry of Trumpler\,37}
\label{sec:cmd}
Photometry of Trumpler\,37 in the 2MASS $J$, $H$ and $K_S$ bands was extracted
from the catalogue II/246 \citep{cutri03} available in the
VizieR\footnote{\it http://vizier.u-strasbg.fr/viz-bin/VizieR} data base
\citep{ocbama00} in a circular area of radius 350\,arcmin centred on HR\,8281
(Table\,\ref{tab:tr37f} -- coordinates from the
SIMBAD\footnote{\it http://simbad.u-strasbg.fr/simbad/} astronomical
data base; \citealt{oberto06}) for a total of 1361458 stars. As quality
constraint, we only considered stars with photometric errors $\leq$\,0.1\,mag in
each band.

\begin{figure*}
\resizebox{0.95\hsize}{!}{\includegraphics{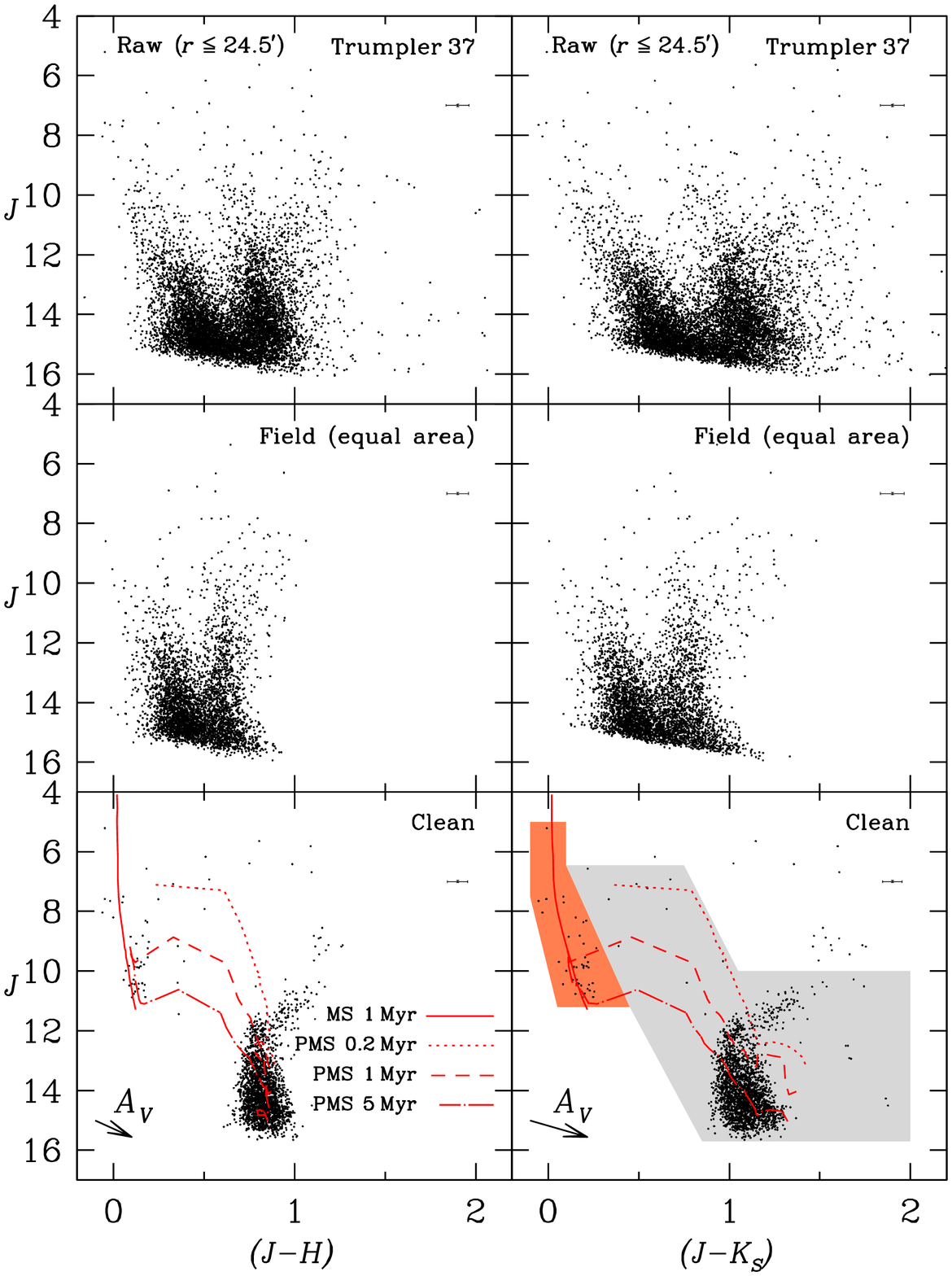}}
\caption[]{2MASS colour-magnitude diagrams extracted from the
{\it r}\,$\leq$\,24.5\,arcmin region of Trumpler\,37.
Mean uncertainties are represented by error bars in the upper right
of each panel. Top panels: observed photometry with {\it J$\times$(J$-$H)}
(left) and {\it J$\times$(J$-$K$_S$)} (right). Middle: equal-area extraction
from the comparison field. Bottom panels: decontaminated CMDs with 1\,Myr
Solar-metallicity Padova isochrone \citep{marigo08} and 0.2, 1 and 5\,Myr
PMS isochrones \citep{siess00}. Heavy-shaded polygon: colour-magnitude
filter to isolate the MS stars. Light-shaded polygon: colour-magnitude filter
for the PMS stars. The arrows indicate the reddening vectors for
$A_V$\,=\,1.8\,mag.}
\label{fig:tr37_cmd}
\end{figure*}

Fig.\,\ref{fig:tr37_cmd} displays {\it J$\times$(J$-$H)} and
{\it J$\times$(J$-$K$_S$)} CMDs for the Trumpler\,37 stellar content and field
stars for an offset equal area. This is simply for comparison purposes, since
the decontamination for field stars was based on a large field within the range
of $r$\,=\,150$-$350\,arcmin, together with an outlying circular area at a
distance of 13.82$^{\circ}$ and a radius of 100\,arcmin centred on
$\alpha$\,=\,21$^{\rm h}$51$^{\rm m}$00$^{\rm s}$ and
$\delta$\,=\,+44$^{\circ}$00$''$00$'$ to improve statistics. The membership of
stars of the cluster was determined by comparison of the photometric properties
of the stars in the offset field and in the cluster (e.g. \citealt{bonbic09}).

The decontamination algorithm works as follows:

\begin{enumerate}\itemsep5pt
\item A spatial extraction of radius 24.5\,arcmin, large enough to be statistically
representative of colours and magnitude of Trumpler\,37, was chosen by eye.
Three-dimensional CMDs $J$$\times${\it (J$-$H)}$\times${\it (J$-$K$_S$)}
were build for this region and for the field. The step along each dimension was
${\it \Delta} J$\,=\,1.0 and ${\it \Delta}
(J$$-$$H)$\,=\,${\it \Delta} (J$$-$$K_S)$\,=\,0.2. For each cluster CMD cell we
modeled the contamination based on the comparison field. Photometric
uncertainties were taken into account, in the sense that what we computed was
the probability of a given star to be found in a cell (this minimizes the
difference of the error function between the borders of the cells). Given the
probabilities of all stars, we obtained the density of stars in each cell. We
carried out this procedure for the object and comparison field cells.

\item Subsequently, the comparison field density was converted back into an
integer number of stars, which was subtracted from the cluster extraction, on a
cell-by-cell basis, resulting the number of member stars in a given cell,
$N^{cell}_{clean}$. Variations in the cell positioning corresponding to 1/3 of
the adopted cell size in each dimension were allowed, such that we ran 243
different setups. Each setup produced a total number of member stars,
$N_{mem}\,=\,\sum_{cell}N^{cell}_{clean}$, from which we computed the expected
total number of member stars $\left<N_{mem}\right>$ by averaging out $N_{mem}$
over all combinations.

\item The stars were ranked according to the number of times they survived the
243 runs, and only the $\left<N_{mem}\right>$ highest ranked stars were
considered cluster members and transposed to the respective decontaminated CMD
(bottom panels of Fig.\,\ref{fig:tr37_cmd}). Only 2267 from the 7972 stars that
were within the circle of radius 24.5\,arcmin have survived the decontamination
procedure. The difference between the expected number of field stars (sometimes
fractional) and the number of stars effectively subtracted (integer) from each
cell is the subtraction efficiency, which summed over all cells resulted
94.6$\pm$0.5\,per cent.
\end{enumerate}

The resulting decontaminated {\it J$\times$(J$-$H)} and
{\it J$\times$(J$-$K$_S$)} CMDs of Trumpler\,37 are shown in
Fig.\,\ref{fig:tr37_cmd}. They reveal a prominent gap between MS and PMS stars
(bottom panels in Fig.\,\ref{fig:tr37_cmd}) even wider than that of the probable
young dissolving cluster vdB\,92 \citep{bonbic10}.

We found that the 1\,Myr Padova main-sequence (MS) isochrone
\citep{marigo08} and the pre-main sequence (PMS) isochrones \citep*{siess00} for
the ages 0.2, 1 and 5\,Myr are the ones that best describe the sequences in the
decontaminated CMDs. Since the 2MASS photometric uncertainties do not allow to
detect metallicity differences, all the adopted isochrones have
Solar-metallicity, suitable for the Galactic disc in general. For the
purposes of this work we adopted 5\,Myr as the age of Trumpler\,37
(Table\,\ref{tab:tr37f}).

\citet{morbid97} estimated the absorption of fourteen bright OB stars of
Trumpler\,37 and suggested that the reddening towards them appears to be
constant and of foreground origin. We used their mean absorption
$A_V$\,=\,1.80$\pm$0.48\,mag and a distance from the Sun
{\it d$_{\odot}$}\,=\,800\,pc (Sect.\,\ref{sec:in}) as constraints to set the
isochrones (Fig.\,\ref{fig:tr37_cmd}). We remark that in such a young cluster,
the star formation process is expected to extend over a time comparable to the
cluster age (e.g. \citealt{stauffer97}). Therefore, it may be very difficult to
disentangle the age spread and the differential reddening, especially when
photometry is the only available information \citep*{bobili11}.
Table\,\ref{tab:tr37f} shows the coordinates and fundamental parameters of
Trumpler\,37.

\begin{figure}
\resizebox{\hsize}{!}{\includegraphics{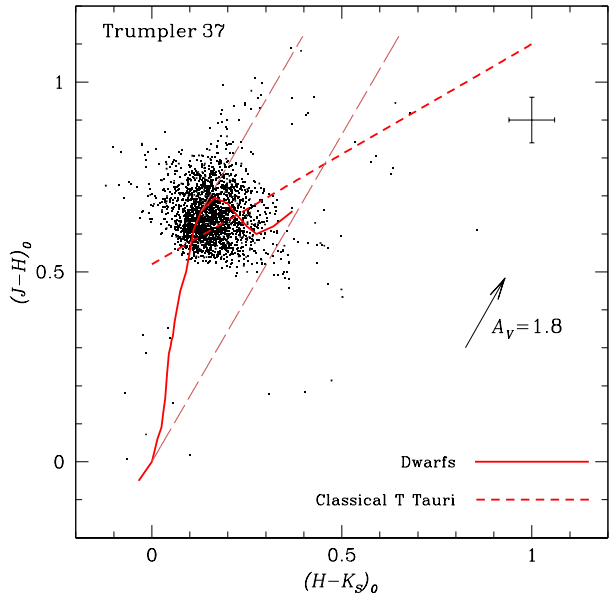}}
\caption[]{Extinction-corrected colour-colour diagrams
{\it (H$-$K$_S$)$\it _0$$\times$(J$-$H)$\it _0$} of the PMS stars of
Trumpler\,37. Mean uncertainties are represented by error bars in upper right.
The solid line is the standard sequence of dwarf stars \citep{besbre88} and
the short-dashed line is the Classical T Tauri stars locus \citep{meyer97}.
The arrow indicates the reddening vector for $A_V$\,=\,1.8\,mag. The two
long-dashed lines define a parallel band to the reddening vector. The stars with
colours that fall outside and to the right of the band have infrared excess
emission, while stars that fall outside and to the left may be residual field
contamination.}
\label{fig:tr37_2cd}
\end{figure}

We applied the relations given by \citet*{dusabi02} to the magnitudes of the PMS
stars of our sample stars using the reddening values of Table\,\ref{tab:tr37f}
in order to build an intrinsic colour-colour diagram
{\it (H$-$K$_S$)$\it _0$$\times$(J$-$H)$\it _0$} of Trumpler\,37
(Fig.\,\ref{fig:tr37_2cd}).
The Classical T Tauri locus \citep*{meyer97} and the standard sequence of
dwarf stars \citep{besbre88} are shown for comparison purposes. Young
stellar objects of different types tend to occupy different regions in the
colour-colour diagram \citep{ladada92}. Two long-dashed lines parallel to
the reddening vector are shown in Fig.\,\ref{fig:tr37_2cd}, such that stars with
colours outside and to the right of the delimiting bands are sources with
infrared excess (a disk emission indicator). Herbig Ae/Be stars in general have
infrared excesses larger than other young stellar objects and are located in a
region delimited by {\it (H$-$K$_S$)$\it _0$}\,$>$\,0.5\,mag and
{\it (J$-$H)$\it _0$}\,$>$\,0.5\,mag.

\begin{figure}
\begin{minipage}[c]{\textwidth}
\includegraphics[width=0.235\textwidth]{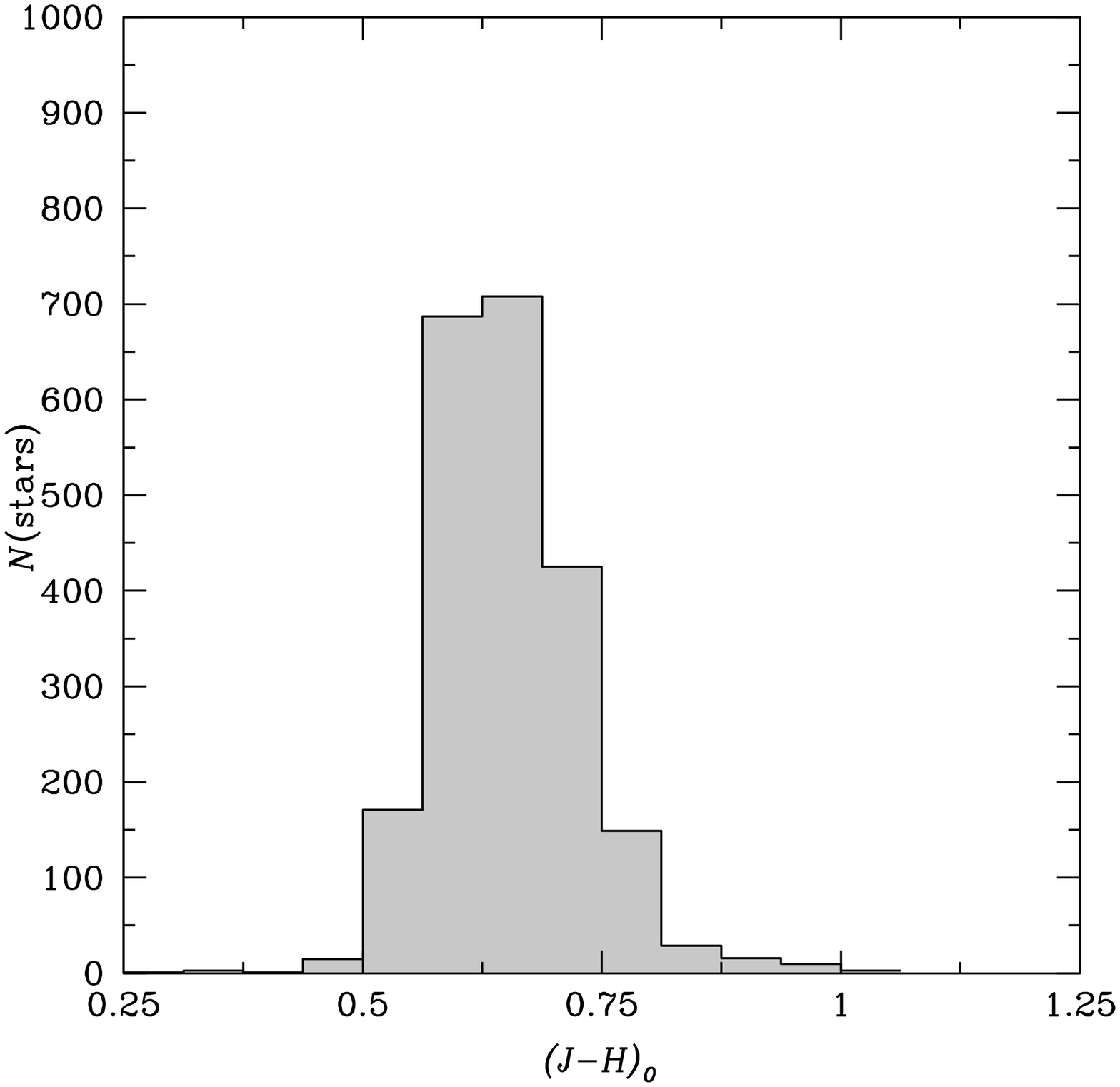}
\includegraphics[width=0.235\textwidth]{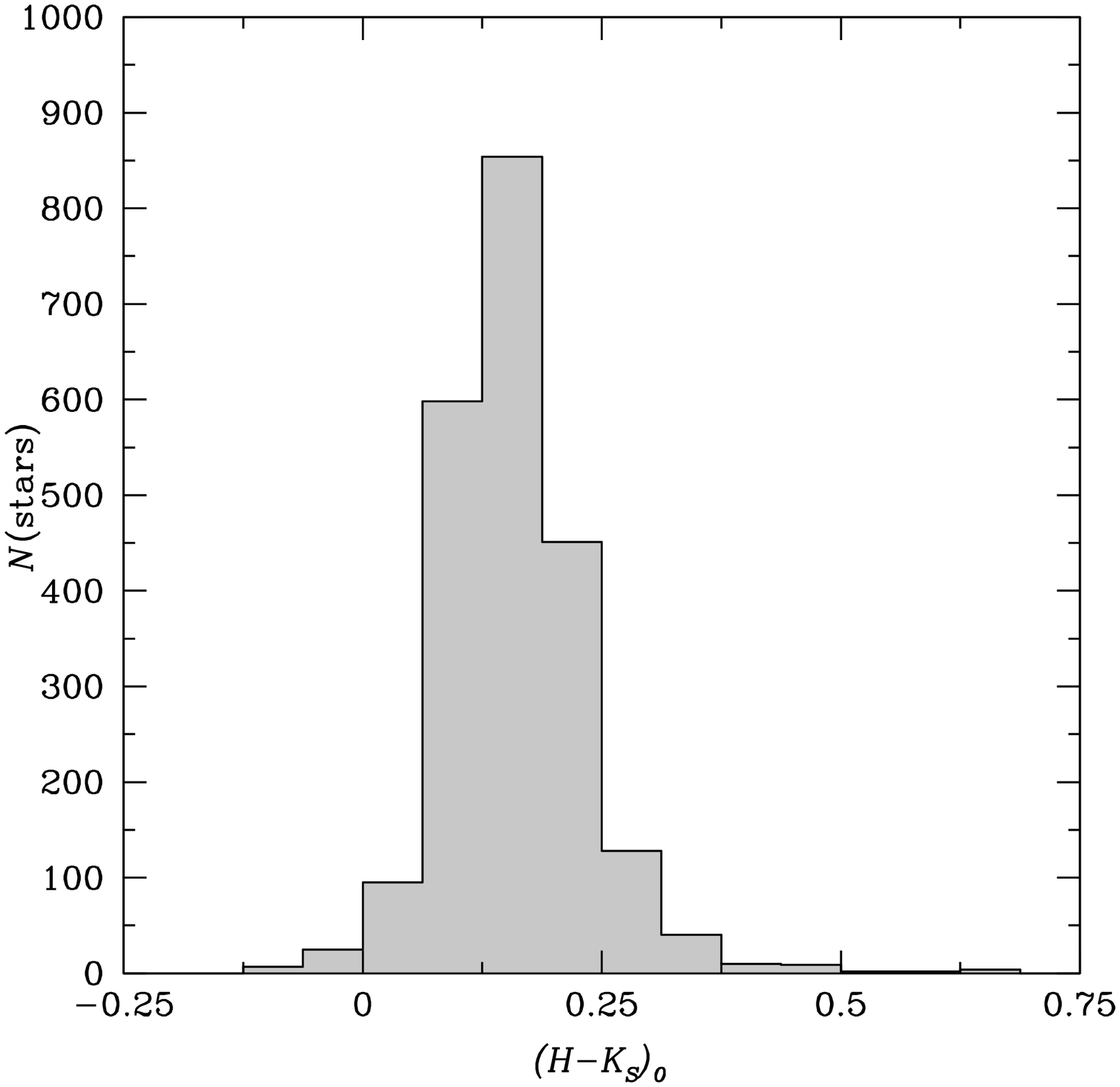}
\end{minipage}
\caption[]{Distributions of the foreground extinction-corrected
($A_V$\,=\,1.8\,mag) colours of the PMS stars of Trumpler\,37.}
\label{fig:tr37_histo}
\end{figure}

Fig.\,\ref{fig:tr37_histo} shows histograms for extinction-corrected colours
{\it (H$-$K$_S$)$\it _0$} and {\it (J$-$H)$\it _0$} of the PMS stars of
Trumpler\,37. About 33\,per cent
of these stars have a {\it (H$-$K$_S$)$\it _0$} excess that can be attributed to
residual field star contamination \citep{froebr05}, but it does not affect
the colour-magnitude filter limits (shaded areas in Fig.\,\ref{fig:tr37_cmd}).
The limits of these filters do not require high precision and were selected
by eye. They are applied to the raw CMD within the range $r$\,=\,0$'$$-$350$'$
to minimise contamination of non-cluster stars before building the stellar RDPs
(Sect.\,\ref{sec:rdp}).

\begin{table*}
\caption[]{Trumpler\,37: fundamental parameters. By columns:
(1) Galactic longitude$^{\dagger}$;
(2) Galactic latitude$^{\dagger}$;
(3) right ascension (J2000)$^{\dagger}$;
(4) declination (J2000)$^{\dagger}$;
(5) distance modulus;
(6) colour excess {\it E(J$-$H)};
(7) colour excess {\it E(J$-$K$_S$)};
(8) colour excess {\it E(B$-$V)};
(9) $V$ band absorption;
(10) adopted age;
(11) distance from the Sun;
(12) Galactocentric distance considering {\it R$_{\odot}$}\,=\,8.4\,kpc \citep{reid09}.
$^{\dagger}$Coordinates from the SIMBAD database.}
\label{tab:tr37f}
\renewcommand{\tabcolsep}{1.3mm}
\renewcommand{\arraystretch}{1.25}
\begin{tabular}{cccccccccccc}
\hline
\multicolumn{1}{c}{$l$} &\multicolumn{1}{c}{$b$}
&\multicolumn{1}{c}{$\alpha$} &\multicolumn{1}{c}{$\delta$}
&\multicolumn{1}{c}{\it (m$-$M)J} &\multicolumn{1}{c}{\it E(J$-$H)}
&\multicolumn{1}{c}{\it E(J$-$K$_S$)} &\multicolumn{1}{c}{\it E(B$-$V)}
&\multicolumn{1}{c}{$A_V$} &\multicolumn{1}{c}{Age}
&\multicolumn{1}{c}{\it d$_{\odot}$} &\multicolumn{1}{c}{\it d$_{GC}$}\\
&&&&&&&&&\multicolumn{1}{c}{(Myr)} &\multicolumn{1}{c}{(kpc)} &\multicolumn{1}{c}{(kpc)}\\
\multicolumn{1}{c}{(1)} &\multicolumn{1}{c}{(2)}
&\multicolumn{1}{c}{(3)} &\multicolumn{1}{c}{(4)}
&\multicolumn{1}{c}{(5)} &\multicolumn{1}{c}{(6)}
&\multicolumn{1}{c}{(7)} &\multicolumn{1}{c}{(8)}
&\multicolumn{1}{c}{(9)} &\multicolumn{1}{c}{(10)}
&\multicolumn{1}{c}{(11)} &\multicolumn{1}{c}{(12)}\\
\hline
99$^{\circ}_{\upf{.}}$29 & 3$^{\circ}_{\upf{.}}$74 & 21$^{\rm h}$38$^{\rm m}$57$^{\rm s}_{\upf{.}}$6 & 57$^{\circ}$29$'$20$''_{\upf{.}}$5 & 10.02$\pm$0.09 & 0.18$\pm$0.05 & 0.28$\pm$0.08 & 0.58$\pm$0.15 & 1.80$\pm$0.48 & 5 & 0.80$\pm$0.06 & 8.57$\pm$0.01\\
\hline
\end{tabular}
\end{table*}

\section{Stellar density profiles}
\label{sec:rdp}
\begin{figure}
\resizebox{\hsize}{!}{\includegraphics{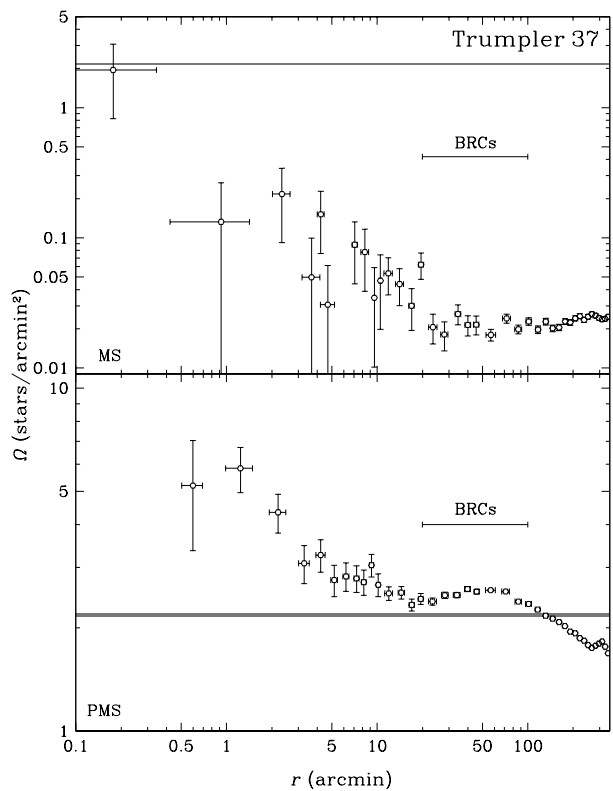}}
\caption[]{Stellar radial density profiles of Trumpler\,37 built separately for
MS (top panel) and PMS (bottom panel) with individual colour-magnitude filters
(Fig.\,\ref{fig:tr37_cmd}). Background stellar level
(Table\,\ref{tab:tr37p}) with 3$\sigma$ is represented by the horizontal
shaded stripes. Since there are few and sparse main-sequence stars, the
corresponding RDP is below the background stellar level. The BRC locus
(Sect.\,\ref{sec:brc}) is indicated and it matches the bump of PMS stars in
20$'$\,$\lesssim$\,$r$\,$\lesssim$\,100$'$.}
\label{fig:tr37_rdp2}
\end{figure}

\begin{figure*}
\resizebox{\hsize}{!}{\includegraphics[angle=270]{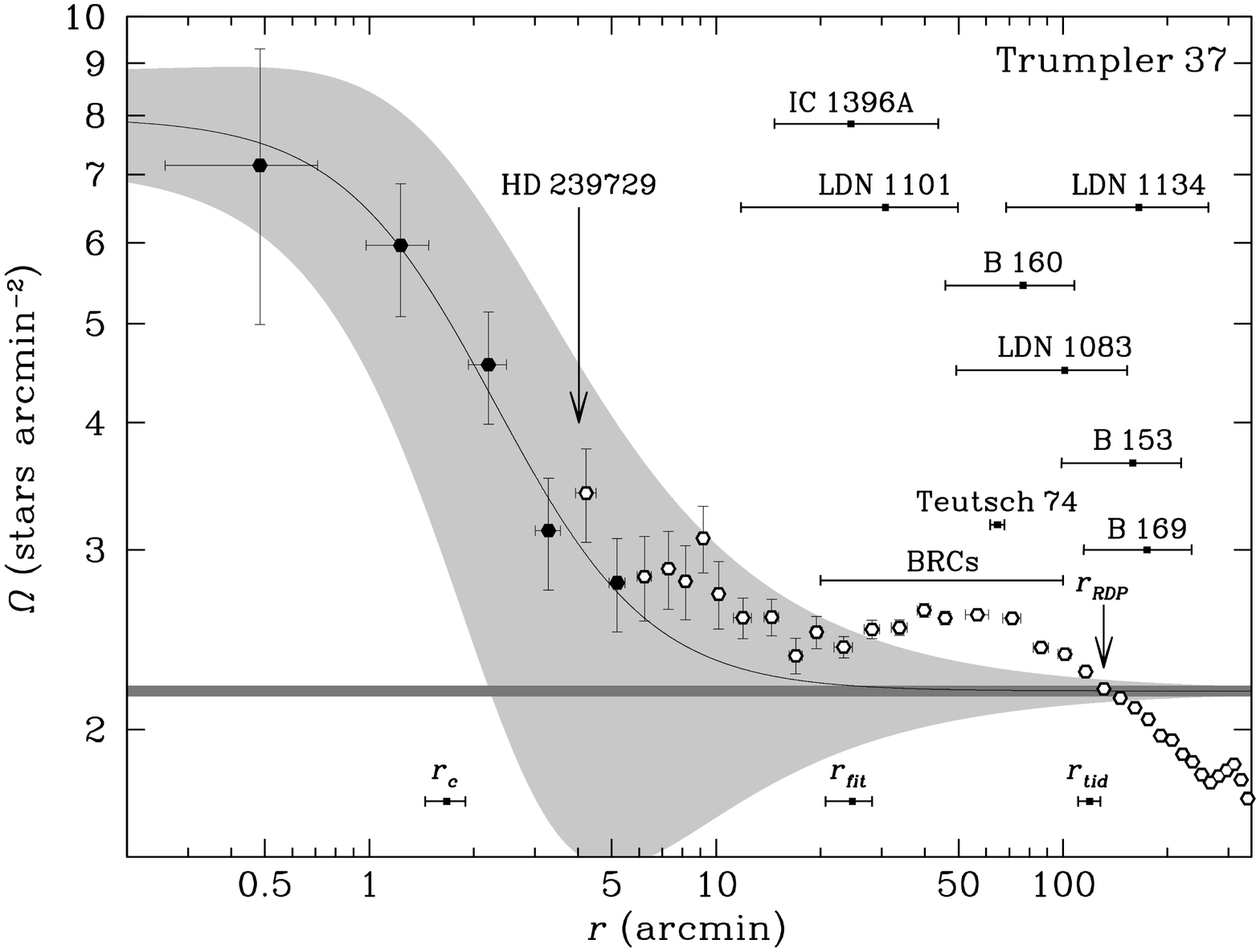}}
\caption[]{Total stellar radial density profile of Trumpler\,37 built with both
colour-magnitude filters (Fig.\,\ref{fig:tr37_cmd}) centred on HR\,8281. The
King-like model (Eq.\,\ref{eq:king}) fitted is shown as a solid line. Open
circles were excluded from the fit. Background stellar level
(Table\,\ref{tab:tr37p}) with 3$\sigma$ is represented by the horizontal
shaded stripe and 1$\sigma$ fit uncertainty is represented by the shaded region
along the fit. The core radius ($r_c$), fitting radius ({\it r$_{fit}$}), tidal
radius ({\it r$_{tid}$}), and RDP radius ({\it r$_{RDP}$}) are shown. These
radii are indicated in Fig.\,\ref{fig:ic1396}. The nearby projected open
cluster Teutsch\,74, the HD\,239729 multiple star, the Elephant Trunk Nebula
(IC\,1396A), BRC locus (Sect.\,\ref{sec:brc}) and some surrounding dark nebulae
are indicated.}
\label{fig:tr37_rdp}
\end{figure*}

Since Trumpler\,37 has very distinct MS and PMS populations, we define two
colour-magnitude filters (shaded polygons in Fig.\,\ref{fig:tr37_cmd}) and build
separate stellar RDPs for each sequence (Fig.\,\ref{fig:tr37_rdp2}). By
comparing these profiles it is possible to see a predominance of PMS stars
throughout the object. A total RDP (MS\,+\,PMS) was also built
(Fig.\,\ref{fig:tr37_rdp}). All these radial profiles were built with concentric
annuli and the x-axis positions correspond to the most populated radial position
inside each annulus.

The positions of other objects projected in the direction of Trumpler\,37 are
also indicated in Fig.\,\ref{fig:tr37_rdp}. There are many dark nebulae in that
region according to the catalogues VII/220A/barnard \citep{barnar27} and the
VII/7A/ldn \citep{lynds62} in the VizieR data base, but for clarity we marked
only some of the larger nebulae in Fig.\,\ref{fig:tr37_rdp}.

In order to derive the structural parameters of Trumpler\,37, a King-like
profile \citep{kin62} given by

\begin{equation}
{\it \Omega} (r) = {\it \Omega_{bg}} + \frac{\it \Omega_{0}}{1 + (r/r_c)^2}
\label{eq:king}
\end{equation}
was fitted to the inner region (filled points in Fig.\,\ref{fig:tr37_rdp}) of
the MS and PMS composite RDP using a non-linear least-squares routine that uses
the y-axis errors as weights. The structural parameters central surface density
(${\it \Omega_{0}}$) and core radius ($r_c$) were obtained from this fit, while
the background density (${\it \Omega_{bg}}$) was measured in a surrounding
annulus in the range $r$\,=\,120$'-$150$'$. The cluster fitting radius
({\it r$_{fit}$}), defined as the projected distance from the cluster centre
where the fitted curve and the background are indistinguishable, was also
obtained. The values of all these parameters are shown in
Tables\,\ref{tab:tr37p}-\ref{tab:tr37r}. Fig.\,\ref{fig:ic1396} shows a
comparison between scales of the core and the fitting radii.

An estimate of the embedded mass ({\it M$_{emb}$}) of Trumpler\,37 was made
considering only the stars on the field-decontaminated CMD
(Fig.\,\ref{fig:tr37_cmd}). The mass of each MS star was determined from
the corresponding mass-luminosity relation of the 1\,Myr Solar-metallicity
Padova isochrone (Sect.\,\ref{sec:cmd}) using dereddened colours and magnitudes
within the range 3.5\,M$_{\odot}-$40\,M$_{\odot}$. Summing the individual
values, we found a total MS mass of 330$^{+10}_{-60}$\,M$_{\odot}$. The
uncertainties take into account reddening, colour excess and individual
stellar mass uncertainties. Given the differential reddening, age spread and the
uncertainties related to the isochrone setting, it is not possible to derive the
mass of each PMS star. Consequently, we simply count the number of PMS stars and
multiply it by a mean PMS stellar mass. For the latter we assumed an initial
mass function of \citet{kroupa01} between 0.08\,M$_{\odot}$ and 7\,M$_{\odot}$,
which results in a mean mass of 0.6\,M$_{\odot}$ with a negligible uncertainty.
Multiplying the number of PMS stars with infrared excess
(Fig.\,\ref{fig:tr37_2cd}) by this mean mass, yields $\sim$890\,M$_{\odot}$.
Therefore, the total embedded mass inside {\it r$_{fit}$} is
1220$^{+10}_{-60}$\,M$_{\odot}$. Note that these values must be considered as
lower limits, owing to (i) the presence of dust, (ii) the fact that we have
considered only the stars of the decontaminated CMD, (iii) the profile deviates
from an isothermal sphere model (Eq.\,\ref{eq:king}) for $r$\,$\gtrsim$\,6$'$,
and (iv) it has a bump beyond $r$\,$\gtrsim$\,20$'$. The measured profile
becomes statistically indistinguishable from the background at
{\it R$_{RDP}$}\,$\approx$\,131\,arcmin (Fig.\,\ref{fig:ic1396}), this can be the
intrinsic extent of Trumpler\,37. In Sect.\,\ref{sec:brc} we analyze this bump
in the profile and reestimate the stellar mass of Trumpler\,37.

\begin{table*}
\caption[]{Trumpler\,37: structural parameters. By columns:
(1) background stellar density;
(2) central stellar density;
(3) angular core radius;
(4) linear core radius.}
\label{tab:tr37p}
\begin{tabular}{cccc}
\hline
\multicolumn{1}{c}{$\it \Omega_{bg}$} &\multicolumn{1}{c}{$\it \Omega_{0}$}
&\multicolumn{1}{c}{$r_c$} &\multicolumn{1}{c}{$R_c$}\\
\multicolumn{1}{c}{(stars arcmin$^{-2}$)}&\multicolumn{1}{c}{(stars arcmin$^{-2}$)}
&\multicolumn{1}{c}{(arcmin)}&\multicolumn{1}{c}{(pc)}\\
\multicolumn{1}{c}{(1)} &\multicolumn{1}{c}{(2)}
&\multicolumn{1}{c}{(3)} &\multicolumn{1}{c}{(4)}\\
\hline
2.18$\pm$0.01 & 5.78$\pm$0.88 & 1.7$\pm$0.2 & 0.4$\pm$0.1\\
\hline
\end{tabular}
\end{table*}

\begin{table*}
\caption[]{Trumpler\,37: scale, mass and crossing time. By columns:
(1) angular fitting radius;
(2) linear fitting radius;
(3) velocity dispersion within the fitting radius;
(4) crossing time of the fitting radius;
(5) total embedded mass within the fitting radius;
(6) angular tidal radius;
(7) linear tidal radius;
(8) angular RDP radius;
(9) linear RDP radius;
(10) velocity dispersion within the tidal radius;
(11) crossing time of the tidal radius;
(12) total embedded mass within the tidal radius;
(13) star formation efficiency.}
\label{tab:tr37r}
\renewcommand{\tabcolsep}{1.825mm}
\renewcommand{\arraystretch}{1.25}
\begin{tabular}{ccccccccccccc}
\hline
\multicolumn{1}{c}{\it r$_{fit}$} &\multicolumn{1}{c}{\it R$_{fit}$}
&\multicolumn{1}{c}{\it M$^{fit}_{emb}$} &\multicolumn{1}{c}{\it $\sigma^{fit}_{V}$}
&\multicolumn{1}{c}{\it $\tau^{fit}_{cr}$}
&\multicolumn{1}{c}{\it r$_{tid}$} &\multicolumn{1}{c}{\it R$_{tid}$}
&\multicolumn{1}{c}{\it r$_{RDP}$} &\multicolumn{1}{c}{\it R$_{RDP}$}
&\multicolumn{1}{c}{\it M$^{tid}_{emb}$} &\multicolumn{1}{c}{\it $\sigma^{tid}_{V}$}
&\multicolumn{1}{c}{\it $\tau^{tid}_{cr}$} &\multicolumn{1}{c}{SFE}\\
\multicolumn{1}{c}{(arcmin)} &\multicolumn{1}{c}{(pc)}
&\multicolumn{1}{c}{(M$_{\odot}$)}
&\multicolumn{1}{c}{km\,s$^{-1}$} &\multicolumn{1}{c}{(Myr)}
&\multicolumn{1}{c}{(arcmin)} &\multicolumn{1}{c}{(pc)}
&\multicolumn{1}{c}{(arcmin)} &\multicolumn{1}{c}{(pc)}
&\multicolumn{1}{c}{(M$_{\odot}$)}
&\multicolumn{1}{c}{km\,s$^{-1}$} &\multicolumn{1}{c}{(Myr)} &\\
\multicolumn{1}{c}{(1)} &\multicolumn{1}{c}{(2)}
&\multicolumn{1}{c}{(3)} &\multicolumn{1}{c}{(4)}
&\multicolumn{1}{c}{(5)} &\multicolumn{1}{c}{(6)}
&\multicolumn{1}{c}{(7)} &\multicolumn{1}{c}{(8)}
&\multicolumn{1}{c}{(9)} &\multicolumn{1}{c}{(10)}
&\multicolumn{1}{c}{(11)} &\multicolumn{1}{c}{(12)} &\multicolumn{1}{c}{(13)}\\
\hline
24.4$\pm$3.7 & 5.7$\pm$1.0 & 1220$^{+10}_{-60}$ & 3.2$\pm$0.3 & 3.5$\pm$2 & 119.4$\pm$8.7 & 27.8$\pm$2.0 & 131 & 31 & 1300$^{+10}_{-60}$ & 1.4$\pm$0.1 & 38$\pm$22 & 0.1\\
\hline
\end{tabular}
\end{table*}

\subsection{Bright-Rimmed Clouds in IC\,1396}
\label{sec:brc}
There are 11 bright-rimmed clouds (BRCs) in IC\,1396 \citep{sufuog91}. All
of them are associated with IRAS (Infrared Astronomical Satellite) sources,
presumably protostars, and may be sites of sequential star formation triggered
by winds and ultra-violet radiation from nearby massive stars, mainly HR\,8281.

\begin{figure*}
\resizebox{0.95\hsize}{!}{\includegraphics{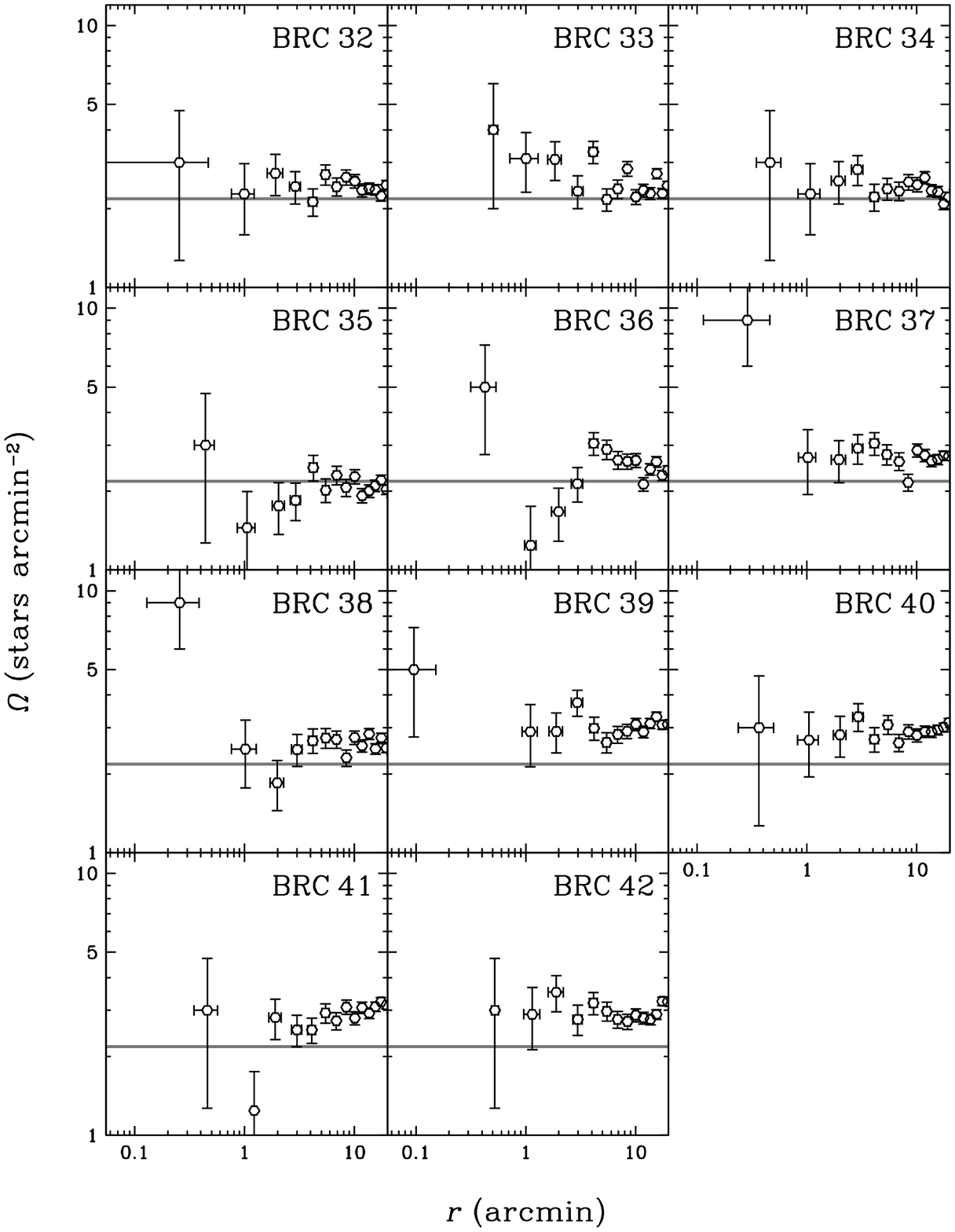}}
\caption[]{Stellar radial density profiles in bright-rimmed clouds in IC\,1396
built with the PMS colour-magnitude filter defined for Trumpler\,37
(light-shaded polygon in Fig.\,\ref{fig:tr37_cmd}). The background stellar level
(Table\,\ref{tab:tr37p}) with 3$\sigma$ is represented by the narrow
horizontal shaded stripes. The objects BRC\,33, BRC\,36, BRC\,37, BRC\,38 and
BRC\,39 have profiles with relatively high central density, suggesting small
star clusters.}
\label{fig:brc_rdp}
\end{figure*}

The location of these BRCs matches the bump for
20$'$\,$\lesssim$\,$r$\,$\lesssim$\,100\,arcmin in
Figs.\,\ref{fig:tr37_rdp2}-\ref{fig:tr37_rdp}.
In order to search for infrared subclusters we built RDPs 
(Fig.\,\ref{fig:brc_rdp}) for each BRC centred in the coordinates listed in
Table\,\ref{tab:brc}. Since that region of IC\,1396 seems to be dominated by
PMS stars (Fig.\,\ref{fig:tr37_rdp2}), we selected only that population
using the PMS colour-magnitude filter defined for Trumpler\,37 (light-shaded
polygon in Fig.\,\ref{fig:tr37_cmd}). The restriction of photometric errors
$\leq$\,0.1\,mag in each band was removed for the stars inside a circle of
radius $r$\,$\sim$\,0.5\,arcmin in BRC\,38 and BRC\,39 due to their high absorption.
An inspection of the RDPs yields BRC\,33, BRC\,36, BRC\,37, BRC\,38 and BRC\,39
as having profiles with relatively high central densities, suggesting small
star clusters. None of them follows a King-like profile (Eq.\,\ref{eq:king}).

\citet{getman07} and \citet{ikeda08} found evidence for the sequential star
formation scenario in BRC\,38 and BRC\,37, respectively. They reported spatial
gradients of stellar age along the direction to the exciting stars and
detected young stellar objects in both BRC areas. The existence of these
subclustering in Trumpler\,37/IC\,1396 resembles the hierarchical structure of
NGC\,346/N\,66 reported by \citet{goulie08} in the Small Magellanic Cloud.

According to \citet{getman07}, BRC\,38 has a stellar mass of
$\sim$15\,M$_{\odot}$. Since BRCs in IC\,1396 have comparable star
counts, we assume the mass of BRC\,38 as representative of such subclusters.
Thus, we estimate that the Trumpler\,37 total embedded mass ({\it M$_{emb}$})
increases to $\sim$1300\,M$_{\odot}$ and its SFE is 10\,per cent.
This is the star formation efficiency for the whole complex. The clumps of the
molecular cloud may have a higher efficiency.

\begin{table*}
\caption[]{Positions determined in this study for the bright-rimmed clouds in IC\,1396. By columns:
(1) identification;
(2) right ascension (J2000);
(3) declination (J2000);
(4) angular projected distance to HR\,8281;
(5) linear projected distance to HR\,8281.}
\label{tab:brc}
\centering
\begin{tabular}{ccccc}
\hline
\multicolumn{1}{c}{BRC} &\multicolumn{1}{c}{$\alpha$} &\multicolumn{1}{c}{$\delta$} &\multicolumn{1}{c}{$d_{HR\,8281}$} &\multicolumn{1}{c}{$D_{HR\,8281}$}\\
&&&\multicolumn{1}{c}{(arcmin)} &\multicolumn{1}{c}{(pc)}\\
\multicolumn{1}{c}{(1)} &\multicolumn{1}{c}{(2)} &\multicolumn{1}{c}{(3)} &\multicolumn{1}{c}{(4)} &\multicolumn{1}{c}{(5)}\\
\hline
32 & 21$^{\rm h}$32$^{\rm m}$34$^{\rm s}_{\upf{.}}$0 & 57$^{\circ}$24$''$27$'$ & 54.1 & 12.6\\
33 & 21$^{\rm h}$33$^{\rm m}$12$^{\rm s}_{\upf{.}}$2 & 57$^{\circ}$29$''$34$'$ & 48.5 & 11.3\\
34 & 21$^{\rm h}$33$^{\rm m}$32$^{\rm s}_{\upf{.}}$4 & 58$^{\circ}$03$''$28$'$ & 57.6 & 13.4\\
35 & 21$^{\rm h}$36$^{\rm m}$05$^{\rm s}_{\upf{.}}$0 & 58$^{\circ}$31$''$09$'$ & 68.7 & 16\\
36 & 21$^{\rm h}$36$^{\rm m}$12$^{\rm s}_{\upf{.}}$4 & 57$^{\circ}$27$''$34$'$ & 23.2 & 5.4\\
37 & 21$^{\rm h}$40$^{\rm m}$27$^{\rm s}_{\upf{.}}$0 & 56$^{\circ}$36$''$16$'$ & 56.7 & 13.2\\
38 & 21$^{\rm h}$40$^{\rm m}$43$^{\rm s}_{\upf{.}}$3 & 58$^{\circ}$15$''$40$'$ & 50.7 & 11.8\\
39 & 21$^{\rm h}$46$^{\rm m}$01$^{\rm s}_{\upf{.}}$5 & 57$^{\circ}$27$''$44$'$ & 59.3 & 13.8\\
40 & 21$^{\rm h}$46$^{\rm m}$12$^{\rm s}_{\upf{.}}$6 & 57$^{\circ}$09$''$59$'$ & 64.4 & 15\\
41 & 21$^{\rm h}$46$^{\rm m}$28$^{\rm s}_{\upf{.}}$6 & 57$^{\circ}$19$''$07$'$ & 64.4 & 15\\
42 & 21$^{\rm h}$46$^{\rm m}$35$^{\rm s}_{\upf{.}}$8 & 57$^{\circ}$12$''$15$'$ & 67   & 15.6\\
\hline
\end{tabular}
\end{table*}

\subsection{Tidal radius and crossing time of Trumpler\,37}
\label{sec:tid}
Since we have an estimate of the total stellar mass in the complex, it is
possible to estimate the dynamical tidal radius ({\it r$_{tid}$}), the distance
from Trumpler\,37 at which stellar motions begin to be affected by the Galaxy,
from the Jacobi limit (Eq.\,7-84 in \citealt{bintre87}) and ignoring any
interaction with surrounding clouds. It can obtained from

\begin{equation}
{\it r_{tid} = \big(\frac{M}{{\rm 3}M_{gal}}\big)^{\rm 1/3} d_{GC},}
\label{eq:tid}
\end{equation}
where {\it M$_{gal}$} is the Galaxy mass within the Galactocentric distance
{\it d$_{GC}$} (Table\,\ref{tab:tr37f}). We derived this mass from the
relation

\begin{equation}
{\it M_{gal} =  \frac{V_{GC}^{\rm 2} d_{GC}}{G},}
\label{eq:mass}
\end{equation}
where $G$ is the gravitational constant, and
{\it V$_{GC}$}\,=\,254\,$\pm$\,16\,km/s is the circular rotation velocity in
{\it R$_{\odot}$}\,=\,8.4$\pm$0.6\,kpc \citep{reid09}. This yields
{\it M$_{gal}$}\,$\approx$\,10$^{11}$\,M$_{\odot}$. Considering
{\it M\,=\,M$_{emb}$+M$_{cld}$}, the tidal radius of Trumpler\,37 results
$\sim$28\,pc ($\sim$119\,arcmin -- see Fig.\,\ref{fig:ic1396}), in good agreement
with the bump end associated to the BRCs.

We can also estimate the crossing time $\tau_{cr}$ by means of

\begin{equation}
{\it \tau_{cr} = \frac{{\rm 2} R}{\sigma_{V}},}
\label{eq:tcr}
\end{equation}
where {\it R} is the cluster radius and $\sigma_{V}$ is the stellar velocity
dispersion given by

\begin{equation}
\it \sigma_{V} = \sqrt{\frac{G M}{R}}.
\label{eq:dis}
\end{equation}
For {\it M\,=\,M$_{emb}$+M$_{cld}$}, with
{\it M$_{emb}$}\,=\,1220\,M$_{\odot}$ and
{\it R}\,=\,{\it R$_{fit}$} (Sect.\,\ref{sec:rdp}), we obtain
$\tau_{cr}$\,$\approx$\,3.5\,Myr, slightly lower than the age of Trumpler\,37.
On the other hand, with {\it M$_{emb}$}\,=\,1300\,M$_{\odot}$ and
{\it R}\,=\,{\it R$_{tid}$}, we obtain $\tau_{cr}$\,$\approx$\,38\,Myr.
Table\,\ref{tab:tr37r} summarizes the estimated parameters.

The comparison of the estimates of the crossing time of Trumpler\,37 with its
age suggests that the inner region ($r$\,$\leq$\,{\it R$_{fit}$}) may remain
bound (Age\,$>$\,$\tau_{cr}$). On the other hand, for the whole
complex ($r$\,$=$\,{\it R$_{tid}$}) we find Age\,$<$\,$\tau_{cr}$, which would
characterize an unbound system. This difference between the two cases possibly
reflects the hierarchical star formation in the molecular cloud and its
fragmentation in the outer regions. This is caused by the radiation of OB
stars and the tidal field of the Galaxy.

\subsection{The background open cluster Teutsch\,74}
\label{sec:teu74}
Teutsch\,74 is an open cluster projected close to IC\,1396
(Fig.\,\ref{fig:ic1396}), so it is necessary to analyze a possible
contamination in the Trumpler\,37 photometry. For this, we used the same tools
as in Sects.\,\ref{sec:cmd}-\ref{sec:rdp}.

Photometry in the 2MASS $J$, $H$ and $K_S$ bands with photometric errors
$\leq$\,0.1\,mag was extracted in a circular area of radius 20\,arcmin centred in the
coordinates given in Table\,\ref{tab:teu74f}.

\begin{figure}
\resizebox{\hsize}{!}{\includegraphics{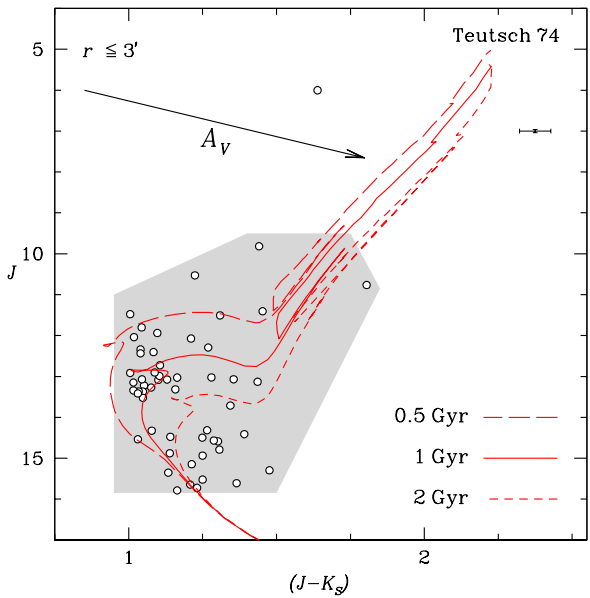}}
\caption[]{Decontaminated colour-magnitude diagram {\it J$\times$(J$-$K$_S$)}
extracted from the {\it r}\,$\leq$\,3\,arcmin region of Teutsch\,74. Mean
uncertainties are represented by error bars in upper right. Solar-metallicity
Padova isochrone \citep{marigo08} for ages of 0.5\,Gyr (long-dashed line),
1\,Gyr (solid line) and 2\,Gyr (short-dashed line) are shown. The shaded
polygon is a colour-magnitude filter. The arrow indicates the reddening vector
for $A_V$\,=\,6.0\,mag.}
\label{fig:teu74_cmd}
\end{figure}

For the field star decontamination we used an annulus with
$r$\,=\,10$'-$20$'$, where the background density (${\it \Omega_{bg}}$)
was measured, and an outlying area of radius
100\,arcmin centred on $\alpha$\,=\,21$^{\rm h}$51$^{\rm m}$00$^{\rm s}$ and
$\delta$\,=\,+44$^{\circ}$00$''$00$'$ (Sect.\,\ref{sec:cmd}). The resulting
subtraction efficiency was 100\,per cent.
Padova isochrones of Solar-metallicity for ages of 0.5\,Gyr, 1\,Gyr and
2\,Gyr \citep{marigo08} have been set by eye to the decontaminated CMD
(Fig.\,\ref{fig:teu74_cmd}) and the fundamental parameters were derived
(Table\,\ref{tab:teu74f}). Teutsch\,74 appears to have a strong
differential reddening, that causes a considerable spread in the CMD.
We assume the age 1\,Gyr of the intermediary isochrone as the approximate
age of Teutsch\,74.

A colour-magnitude filter was defined (shaded area in Fig.\,\ref{fig:teu74_cmd})
and applied to the raw CMD in order to build a stellar RDP with concentric
annuli. The tidal radius and mass were estimated (Table\,\ref{tab:teu74p}).

We subtracted from the Trumpler\,37 photometry all the stars remaining after
decontamination of Teutsch\,74 (Fig.\,\ref{fig:teu74_cmd}) and did not find any
significant change in the results of the Trumpler\,37 analysis.

\begin{figure}
\resizebox{\hsize}{!}{\includegraphics{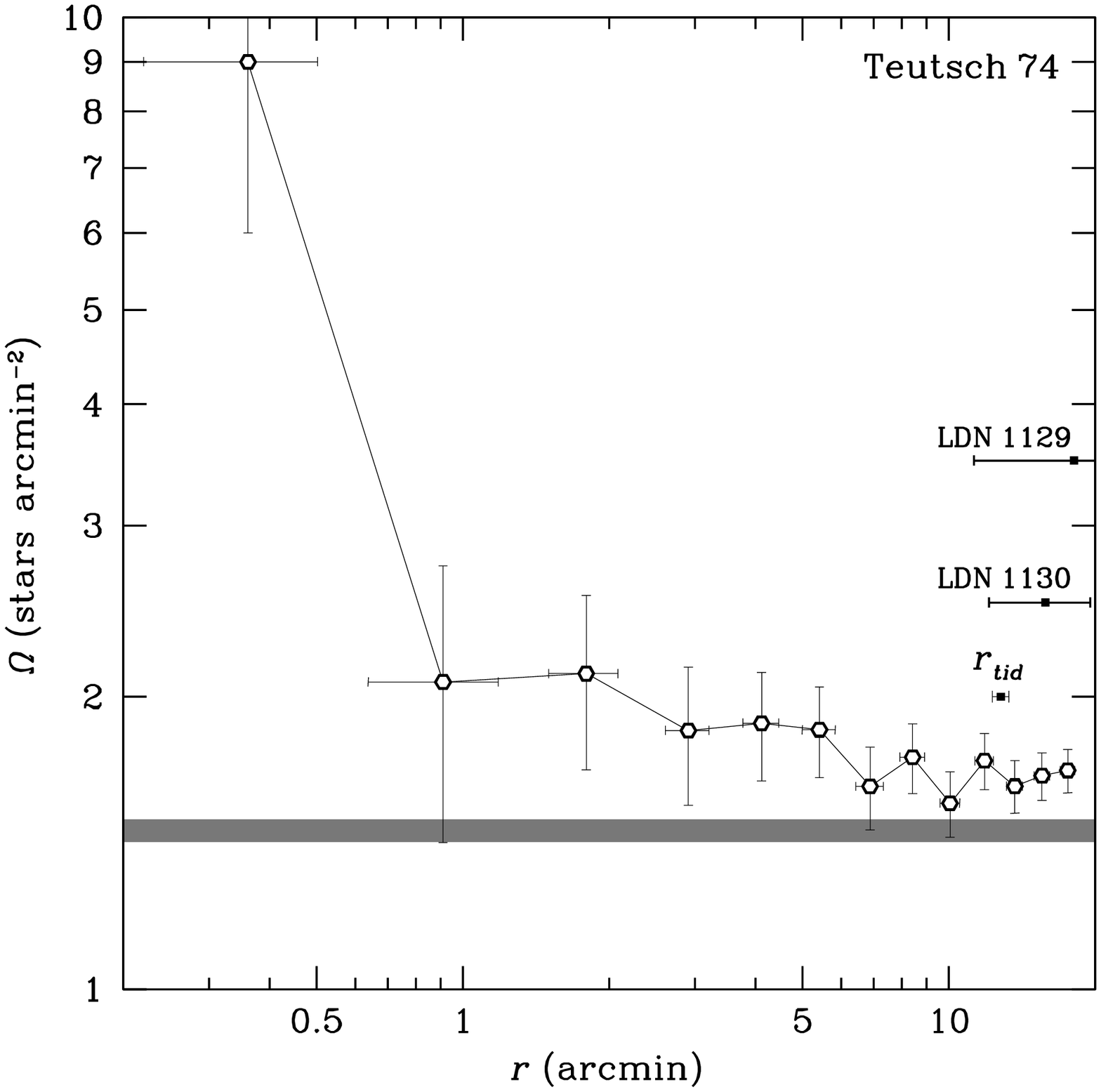}}
\caption[]{Stellar radial density profile of Teutsch\,74 built with the
colour-magnitude filter (Fig.\,\ref{fig:teu74_cmd}). The horizontal shaded
stripe is the background stellar level with 1$\sigma$
(Table\,\ref{tab:teu74p}). Two dark nebulae projected in the direction of
the cluster and the tidal radius ({\it r$_{tid}$} estimated with
Eq.\,\ref{eq:tid}) are indicated.}
\label{fig:teu74_rdp}
\end{figure}

Finally, Teutsch\,74 has a stellar radial density profile
(Fig.\,\ref{fig:teu74_rdp}) that cannot fitted by a King-like model.

\begin{table*}
\caption[]{Teutsch\,74: fundamental parameters. By columns:
(1) Galactic longitude;
(2) Galactic latitude;
(3) right ascension (J2000);
(4) declination (J2000);
(5) distance modulus;
(6) colour excess {\it E(J$-$H)};
(7) colour excess {\it E(J$-$K$_S$)};
(8) colour excess {\it E(B$-$V)};
(9) $V$ band absorption;
(10) adopted age;
(11) distance from the Sun;
(12) Galactocentric distance considering {\it R$_{\odot}$}\,=\,8.4\,kpc \citep{reid09}.}
\label{tab:teu74f}
\renewcommand{\tabcolsep}{1.45mm}
\renewcommand{\arraystretch}{1.25}
\begin{tabular}{cccccccccccc}
\hline
\multicolumn{1}{c}{$l$} &\multicolumn{1}{c}{$b$}
&\multicolumn{1}{c}{$\alpha$} &\multicolumn{1}{c}{$\delta$}
&\multicolumn{1}{c}{\it (m$-$M)J} &\multicolumn{1}{c}{\it E(J$-$H)}
&\multicolumn{1}{c}{\it E(J$-$K$_S$)} &\multicolumn{1}{c}{\it E(B$-$V)}
&\multicolumn{1}{c}{$A_V$} &\multicolumn{1}{c}{Age}
&\multicolumn{1}{c}{\it d$_{\odot}$}
&\multicolumn{1}{c}{\it d$_{GC}$}\\
&&&&&&&&&\multicolumn{1}{c}{(Gyr)} &\multicolumn{1}{c}{(kpc)} &\multicolumn{1}{c}{(kpc)}\\
\multicolumn{1}{c}{(1)} &\multicolumn{1}{c}{(2)}
&\multicolumn{1}{c}{(3)} &\multicolumn{1}{c}{(4)}
&\multicolumn{1}{c}{(5)} &\multicolumn{1}{c}{(6)}
&\multicolumn{1}{c}{(7)} &\multicolumn{1}{c}{(8)}
&\multicolumn{1}{c}{(9)} &\multicolumn{1}{c}{(10)}
&\multicolumn{1}{c}{(11)} &\multicolumn{1}{c}{(12)}\\
\hline
100$^{\circ}_{\upf{.}}$36 & 3$^{\circ}_{\upf{.}}$61 & 21$^{\rm h}$45$^{\rm m}$40$^{\rm s}$ & 58$^{\circ}$05$'$37$''$ & 12.55$\pm$0.70 & 0.61$\pm$0.13 & 0.95$\pm$0.20 & 1.94$\pm$0.40 & 6.00$\pm$1.24 & 1 & 1.50$\pm$0.54 & 8.79$\pm$0.13\\
\hline
\end{tabular}
\end{table*}

\begin{table*}
\caption[]{Teutsch\,74: scale, mass and crossing time. By columns:
(1) background stellar density;
(2) angular tidal radius;
(3) linear tidal radius;
(4) total mass inside the radius {\it r}\,=\,3$'$;
(5) velocity dispersion within the tidal radius;
(6) crossing time of the tidal radius.}
\label{tab:teu74p}
\centering
\begin{tabular}{cccccc}
\hline
\multicolumn{1}{c}{$\it \Omega_{bg}$}
&\multicolumn{1}{c}{\it r$_{tid}$} &\multicolumn{1}{c}{\it R$_{tid}$}
&\multicolumn{1}{c}{\it M} &\multicolumn{1}{c}{\it $\sigma^{tid}_{V}$}
&\multicolumn{1}{c}{\it $\tau_{cr}^{tid}$}\\
\multicolumn{1}{c}{(stars arcmin$^{-2}$)}
&\multicolumn{1}{c}{(arcmin)}&\multicolumn{1}{c}{(pc)}
&\multicolumn{1}{c}{(M$_{\odot}$)} &\multicolumn{1}{c}{(km\,s$^{-1}$)}
&\multicolumn{1}{c}{(Myr)}\\
\multicolumn{1}{c}{(1)} &\multicolumn{1}{c}{(2)}
&\multicolumn{1}{c}{(3)} &\multicolumn{1}{c}{(4)}
&\multicolumn{1}{c}{(5)} &\multicolumn{1}{c}{(6)}\\
\hline
1.46$\pm$0.04 & 12.8$\pm$0.5 & 5.6$\pm$0.2 & 100$\pm$20 & $\sim$0.3 & 39$\pm$7\\
\hline
\end{tabular}
\end{table*}

\section{Discussion}
\label{sec:dis}
\begin{figure*}
\resizebox{0.95\hsize}{!}{\includegraphics{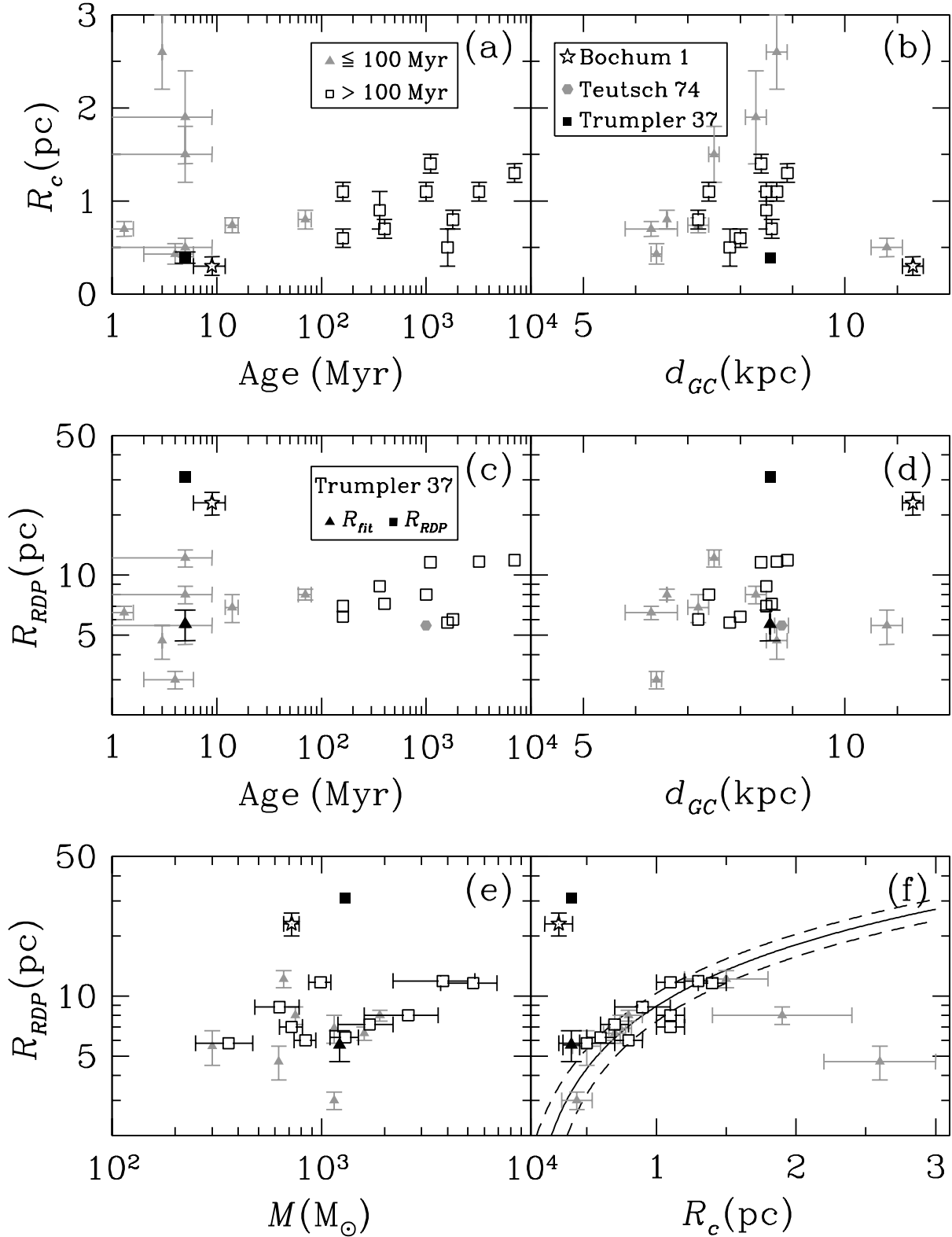}}
\caption[]{Diagrams comparing cluster parameters. Trumpler\,37 is represented as
a filled black square. When we assume {\it R$_{RDP}$}\,$=$\,{\it R$_{fit}$}, a
filled black triangle represents Trumpler\,37. Young clusters ($\leq$100\,Myr)
are represented as gray triangles and old clusters ($>$100\,Myr) as open
squares. Some error bars are smaller than the symbols. The open star represents
Bochum\,1, and the filled gray hexagon represents Teutsch\,74. Solid line in
panel (f) is a fitted curve, and dashed lines are the fitting errors.
Points corresponding to Trumpler\,37, Bochum\,1, vdB\,92 and NGC\,2244 were
not included in the fit.}
\label{fig:tr37_dia}
\end{figure*}

In order to locate Trumpler\,37 in an evolutionary sequence of star clusters we
built diagnostic diagrams (Fig.\,\ref{fig:tr37_dia}) comparing its parameters to
those of objects analyzed in previous studies. The comparison sample consists
of:

\begin{itemize}\itemsep5pt
\item the structurally normal open clusters within a wide age range (70\,Myr to
7\,Gyr) M\,26, NGC\,2287, M\,48, M\,93, NGC\,5822, NGC\,3680, IC\,4651, M\,67
and NGC\,188, plus the populous open clusters NGC\,2477 and NGC\,2516
\citep{bonbic05};
\item NGC\,4755, with residual stellar infrared excess emission and an age of
14\,Myr \citep{bonatt06};
\item the very young and dynamically evolved cluster NGC\,6611, with an age of
1.3\,Myr \citep*{bosabi06};
\item Bochum\,1, with an age of 9\,Myr, a star cluster remnant that might be
evolving into an OB association, and the bound cluster NGC\,6823 with an age of
4\,Myr \citep*{bibodu08};
\item the dissolving young cluster NGC\,2244, with an age of 3\,Myr and a
candidate to ordinary young open cluster NGC\,2239 with an age of 5\,Myr
\citep{bonbic09};
\item the dissolving clusters Collinder\,197 and vdB\,92, both with an age of
5\,Myr \citep{bonbic10};
\item the open cluster Teutsch\,74 (Sect.\,\ref{sec:teu74});
\item and Trumpler\,37 itself.
\end{itemize}

Fig.\,\ref{fig:tr37_dia}, panels (a) and (b) show the dependence of the core
radius on age and Galactocentric distance. Panels (c) and (d) do the same for
the RDP radius. Finally, panels (e) and (f) show the dependence of the RDP
radius on mass and core radius. Young clusters ($\leq$100\,Myr) and old clusters
($>$100\,Myr) are represented by different symbols.

Although the inner region of Trumpler\,37 has been fitted by an isothermal model
(Sect.\,\ref{sec:rdp}), the object does not present typical parameters of a
young star cluster. It has a RDP radius much larger than the fitting radius
(both are represented in Fig.\,\ref{fig:ic1396} and
Fig.\,\ref{fig:tr37_dia}), which is larger than Bochum\,1, an association that
is somewhat older and more distant from the Solar circle. Thus, the extended
nature of Trumpler\,37 can be attributed to tidal effects and molecular cloud
fragmentation. Evidence of the former process is that
{\it R$_{RDP}$}\,$\approx$\,{\it R$_{tid}$}, and of the latter is that there
are many BRCs in the area of Trumpler\,37. Note that the RDP radius might match
the tidal radius as long as Trumpler\,37 is closer to the Sun ($\sim$730\,pc)
or has a larger mass ($\sim$5600\,M$_{\odot}$).

Panel (b) suggests a weak relation of the core radii of the clusters with the
Galactocentric distances, but with a significant spread. In addition, panel (f)
shows a scale relation between the RDP and core radii given by the equation
{\it R$_{RDP}$}\,=\,$(9.16\pm1.06) {\it R_c} - (0.26\pm0.97)$, fitted with a
least-squares routine with errors in both coordinates. Most objects follow this
relation, except Trumpler\,37, Bochum\,1, vdB\,92 and NGC\,2244 which were
not included in the fit. All of them present evidence of dissolving systems. On
the other hand, the inner region of Trumpler\,37, delimited by the fitting
radius, seems to follow this relation, so it might survive as a bound core
of an expanding association.

\section{Concluding Remarks}
\label{sec:con}
In the present paper we analyzed Trumpler\,37 which is embedded in the H\,II
region complex IC\,1396. We used field-star decontaminated 2MASS photometry to
build colour-magnitude diagrams and stellar radial density profiles in order to
find fundamental and structural parameters. The CMD of Trumpler\,37 shows
a detached PMS suggesting sequential star formation and its RDP shows a core
that follows a King-like model surrounded by several small subclusters with
some deeply embedded stars. Trumpler\,37 has a low star formation efficiency
($\sim$10\,per cent)
and Age/$\tau_{cr}$\,$<$\,1, typical of objects that do not survive as bound
systems. In fact, it has parameters similar to those of the objects classified
as dissolving systems in previous studies. Note that these are values for the
whole complex. The star formation efficiency can be higher in the clumps of the
molecular cloud, and Age/$\tau_{cr}$\,$>$\,1 in the inner region of Trumpler\,37
which is fitted by an isothermal sphere model. Anyway, Trumpler\,37 might
characterize an intermediate case between an embedded cluster and an OB
association.

\section*{Acknowledgments}
We thank an anonymous referee for interesting remarks.
We acknowledge support from the Brazilian Institution CNPq.
We acknowledge use of the SIMBAD database and the VizieR Catalogue Service
operated at the CDS, Strasbourg, France.
This publication makes use of data products from the Two Micron All Sky Survey,
which is a joint project of the University of Massachusetts and the
Infrared Processing and Analysis Centre/California Institute of Technology,
funded by the National Aeronautics and Space Administration and the
National Science Foundation.
This research used the facilities of the Canadian Astronomy Data Centre
operated by the National Research Council of Canada with the support of the
Canadian Space Agency.

\label{lastpage}

\end{document}